\documentclass[%
 reprint,
superscriptaddress,
 amsmath,amssymb,
 aps,
]{revtex4-2}

\usepackage{graphicx}% Include figure files
\usepackage{dcolumn}% Align table columns on decimal point
\usepackage{bm}% bold math

\usepackage{siunitx}
\usepackage[utf8]{inputenc}
\DeclareUnicodeCharacter{2212}{\ensuremath{-}}
\begin{document}

\preprint{APS/123-QED}

\title{Transient gap generation in BaFe$_2$As$_2$ driven by coherent lattice vibrations}% Force line breaks with \\

\author{Jacob A. Warshauer}
\author{Daniel Alejandro Bustamante Lopez}
\affiliation{Department of Physics, Boston University, 590 Commonwealth Avenue, Boston, MA 02215, USA}

\author{Qingxin Dong}
\author{Genfu Chen}
\affiliation{Institute of Physics and Beijing National Laboratory for Condensed Matter Physics, Chinese Academy of Sciences, Beijing, 100190, China}
\affiliation{ School of Physical Sciences, University of Chinese Academy of Sciences, Beijing, 100049, China}
\author{Wanzheng Hu}
\email{wanzheng@bu.edu}
\affiliation{Department of Physics, Boston University, 590 Commonwealth Avenue, Boston, MA 02215, USA}
\affiliation{Division of Materials Science and Engineering, Boston University, 590 Commonwealth Avenue, Boston, MA 02215, USA}

%\date{\today}% It is always \today, today,
             %  but any date may be explicitly specified

\begin{abstract}
The electronic structure and the magnetic properties of iron-based superconductors are highly sensitive to the pnictogen height. Coherent excitation of the $A_{1g}$ phonon by femtosecond laser directly modulates the pnictogen height, which has been used to control the physical properties of iron-based superconductors. Previous studies show that the driven $A_{1g}$ phonon resulted in a transient increase of the pnictogen height in BaFe$_2$As$_2$, favoring an enhanced Fe magnetic moment. Here, we use time-resolved broadband terahertz spectroscopy to investigate the dynamics of BaFe$_2$As$_2$ in the $A_{1g}$ phonon driven state. Below the spin-density wave (SDW) transition temperature, we observe a transient gap generation at early time delays. A similar transient feature is observed in the normal state up to room temperature.         
\end{abstract}

%\keywords{Suggested keywords}%Use showkeys class option if keyword
                              %display desired
\maketitle

%\tableofcontents

Controlling the physical properties of quantum materials along non-invasive and ultrafast pathways is the key for developing next-generation technologies. Dynamical control of materials using ultra-short laser pulses has been successful in a variety of systems to achieve novel phases that are inaccessible at equilibrium\cite{zhang2016,stojchevska2014,Li2019,Disa2021}. The key is to identify tuning parameters which effectively modify the electronic structure of quantum materials.

\begin{figure*} 
\includegraphics[scale = 0.6]{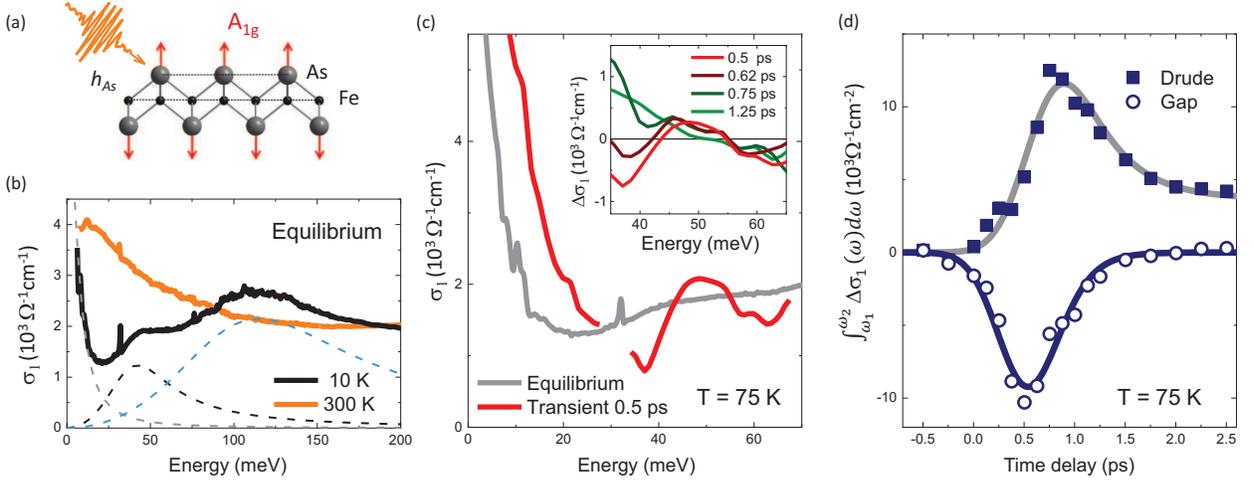}
\caption{\label{fig:s1Ds1} Equilibrium and transient optical response for BaFe$_2$As$_2$ in the SDW state. (a) Coherent excitation of the $A_{1g}$ phonon modulates the pnictogen height $h_{As}$. (b) 
Optical evidence of SDW gap formation at equilibrium: $\sigma_1(\omega)$ at $T = \SI{10}{\kelvin}$ shows peaks at \SI{45}{\milli\eV} and \SI{110}{\milli\eV}, representing the SDW gaps\cite{Hu2008}. Solid lines are $\sigma_1(\omega)$ at temperatures below and above $T_{SDW}$. Dashed lines are the Drude term and two Lorentz peaks from a fit for $T = \SI{10}{\kelvin}$. (c) (Main) Transient optical conductivity (red) at $T = \SI{75}{\kelvin}$. The transient $\sigma_1(\omega)$ shows a broadened Drude component and a gap formation near \SI{50}{\milli\eV} which takes the spectral weight from the equilibrium SDW gap at \SI{45}{\milli\eV}. The data discontinuity at \SI{30}{\milli\eV} is due to a detection gap in the pump-probe setup. The excitation fluence is \SI{0.53}{\milli\joule\per\square\centi\metre}. (Inset) Light-induced change in the real part of optical conductivity, $\Delta\sigma_1(\omega)=\sigma_1(\omega)^{\textrm{transient}}-\sigma_1(\omega)^{\textrm{equilibrium}}$, at early time delays. The transient gap reaches its maximum earlier than the transient Drude component. (d) Different time scales for the light-induced Drude component (solid squares) and the transient gap (open circles) are further evidenced by the time evolution of the transient change in spectral weight, $\int_{\omega_1}^{\omega_2}\Delta\sigma_1(\omega)d\omega$, where $\omega_1$ = \SI{34}{\milli\eV} and $\omega_2$ = \SI{43}{\milli\eV}. Solid lines are exponential fits.}
\end{figure*}

Quantum materials are remarkably sensitive to structural distortion. In iron pnictides, the iron-arsenic distance (pnictogen height) has a significant impact on superconductivity\cite{Lee2008,Kuroki2009}, the electronic band structure\cite{Rettig2012, Avigo2013,Yang2014,Gerber2017} and the magnetic properties\cite{Yin2011,Zhang2014}. The pnictogen height can be periodically modulated by optical excitation of a Raman-active $A_{1g}$ phonon (Fig. \ref{fig:s1Ds1}a)\cite{Litvinchuk2009,Choi2008}. For BaFe$_2$As$_2$, a displacive excitation towards larger pnictogen height is observed in the transient state\cite{Rettig2015,Gerber2015,Yang2014}, which favors an enhanced Fe magnetic moment. One would expect to see a displacive increase of the SDW gap size, but this has not been observed so far due to experimental challenges. In fact, a comprehensive band assignment and gap identification at equilibrium are already challenging\cite{Yi2011, Chauviere2011, Chen2010,Yin2011NP} due to the multi-band nature of iron pnictides: the energy bands close to the Fermi energy come from three orbitals, which experience strong band hybridization and band splitting below the nematic ordering temperature. This situation is further complicated by the twinned domains in as-grown samples\cite{Yi2014,Pfau2019}. In the transient state, the $A_{1g}$ phonon is excited by a femtosecond laser in the near infrared, the energy scale of which is far above that of the SDW gap. This results in a large contribution from the photoexcited carriers in the phonon driven state, which may wash out low-energy features such as the SDW gap. A remarkable result of the transient modification of the SDW gap is from a recent time-resolved optical spectroscopy study, reporting that the averaged oscillation amplitude of the transient optical conductivity $\Delta\sigma_1^{\textrm{osc}}$ resembles the equilibrium gapped optical conductivity at a temperature slightly above the SDW transition temperature $T_{SDW}$\cite{Kim2012}. However, it remains unknown whether a corresponding behavior exists in the displacive response of the SDW gap to counter the enhanced pnictogen height in the phonon driven state\cite{Rettig2015,Gerber2015}.

Here, we report a time-resolved broadband terahertz (THz) spectroscopic probe of BaFe$_2$As$_2$ in the $A_{1g}$ phonon driven state. The THz probe covers a spectral range from 8 to \SI{70}{\milli\eV}, which allows the detection of light-induced changes in the itinerant carriers and the low-energy SDW gap. We studied the time and temperature dependence of the transient optical conductivity $\sigma_1(\omega)$.
We observed a light-induced depletion of the optical conductivity at the equilibrium SDW gap, with a peak forming at higher energies. This is a clear optical signature of gap opening. This feature is observed at early pump-probe time delays, when the photoexcited carriers are accumulating. The transient gap is quickly filled and merges to a free carrier response at later delays. Temperature dependent pump-probe measurements show that a similar transient gap develops above $T_{SDW}$ and persists up to room temperature.

The BaFe$_2$As$_2$ single crystals exhibiting a SDW transition at $T_{SDW} = \SI{130}{\kelvin}$ were grown by self-flux method\cite{Nakajima2009}. Typical dimensions of the as-grown single-crystals are $\sim\SI{3}{\milli\meter}\times\SI{3}{\milli\meter}\times \SI{0.05}{\milli\meter}$. Near-infrared (\SI{800}{\nano\meter}) laser pulses with a \SI{0.53}{\milli\joule\per\square\centi\metre} fluence and \SI{40}{\femto\second} duration were used to excite the $A_{1g}$ phonon in BaFe$_2$As$_2$. The transient optical properties were probed at normal incidence by broadband THz pulses generated by laser-ionized plasma\cite{Ho2010}. The THz pulses were detected by electrooptical (EO) sampling of the terahertz field in a \SI{100}{\micro\meter} thick GaP and a \SI{300}{\micro\meter} thick GaSe crystal to cover a detection range from 8 to \SI{70}{\milli\eV}. A long-pass filter was used to remove the scattered pump photons. For each pump-probe time delay, the relative delay between the pump and the EO sampling pulses were kept fixed while scanning the terahertz transient. This ensures that each point in the terahertz probe field detects the material at the same pump-probe time delay. The terahertz probe field and the pump-induced change in the probe field were simultaneously recorded using two lock-in amplifiers\cite{Iwaszczuk2009}. The complex reflection coefficient of the photoexcited sample was calculated using a multilayer model\cite{Hu2014}, which models the material as a fully photoexcited top layer of a thickness equal to the near-infrared pump penetration depth (\SI{27}{\nano\meter}), and an unexcited bottom layer retaining the equilibrium optical response\cite{Hu2008}. The probe penetration depth is frequency dependent and in the order of \SI{200}{\nano\meter}. 

\begin{figure*}
\includegraphics[scale = 1.0]{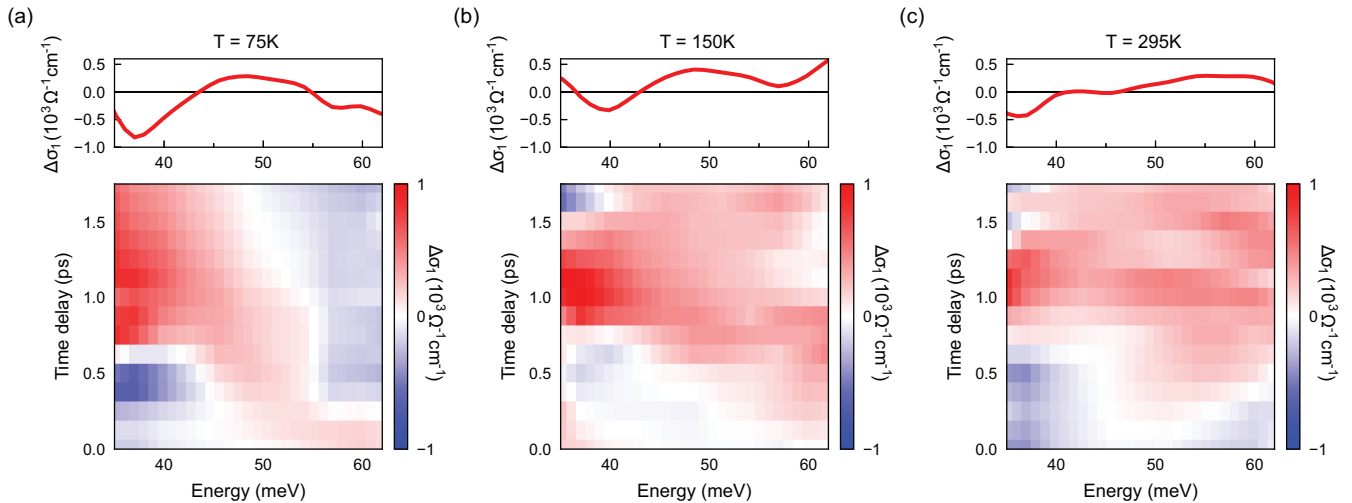}
\caption{\label{fig:Ds1Tdep} Temperature evolution of the transient gap with a pump fluence of \SI{0.53}{\milli\joule\per\square\centi\metre}. (a) $\Delta\sigma_1(\omega)$ in the SDW state at $T = \SI{75}{\kelvin}$. (b) $\Delta\sigma_1(\omega)$ in the normal state at $T = \SI{150}{\kelvin}$. (c) $\Delta\sigma_1(\omega)$ at room temperature. The color plots are $\Delta\sigma_1(\omega)$ at early pump-probe delays; the top panels show $\Delta\sigma_1(\omega)$ at $t$ = \SI{0.5}{\pico\second}. The transient gap generation is seen from below to above the SDW transition temperature.}
\end{figure*}

The equilibrium optical conductivity $\sigma_1(\omega)$, above and below $T_{SDW}$, is shown in Fig. \ref{fig:s1Ds1}(b). In the SDW state, several features are seen in $\sigma_1(\omega)$: a Drude term at low frequencies, representing the free-carrier response; a sharp peak at \SI{32}{\milli\eV} from an $E_u$ infrared-active phonon\cite{Hu2008,Homes2018}; and two peaks at \SI{45}{\milli\eV} and \SI{110}{\milli\eV}, representing the SDW gaps\cite{Hu2008, Chauviere2011,Yin2011NP}.

Figure \ref{fig:s1Ds1}(c) shows the transient $\sigma_1(\omega)$ at $T = \SI{75}{\kelvin}$. At a time delay of \SI{0.5}{\pico\second}, $\sigma_1(\omega)$ shows a broadening of the low-frequency Drude component with an enhanced spectral weight. This indicates a higher carrier scattering rate and an increased carrier density, which are from photoexcited carriers since BaFe$_2$As$_2$ is a metal. In the high-frequency region, the transient $\sigma_1(\omega)$ is suppressed at around \SI{40}{\milli\eV}, and then goes above the equilibrium $\sigma_1(\omega)$ and peaks at \SI{50}{\milli\eV}. This is an optical fingerprint of gap opening. 

With increasing pump-probe time delay, the low-frequency Drude broadening and the high-frequency gap opening evolve with different time scales. In the following, we will focus on the high-frequency region which probes both the transient gap and the tail of the Drude component. The inset of Fig. \ref{fig:s1Ds1}(c) shows the light-induced change in the optical conductivity at selected time delays. The transient gap formation is maximized at \SI{0.5}{\pico\second}. With increasing pump-probe delay, a transient Drude contribution develops, which brings $\Delta\sigma_1(\omega)$ to positive values at \SI{40}{\milli\eV} and weakens the transient gap feature. At \SI{1.25}{\pico\second}, the transient gap disappears, and $\Delta\sigma_1(\omega)$ contains only a Drude term, which decays with time. 

To extract the time evolution of the transient gap and the Drude component, we integrated the light-induced change in the real part of optical conductivity $\int_{\omega_1}^{\omega_2}\Delta\sigma_1(\omega)d\omega$ over the frequency region in which the transient depletion of $\sigma_1(\omega)$ occurs: $\omega_1$ = \SI{34}{\milli\eV} to $\omega_2$ = \SI{43}{\milli\eV}. By fitting the transient spectral weight (see Supplementary Information), we separated the contributions of the transient gap and the transient Drude component. In Fig. \ref{fig:s1Ds1}(d), the open circles represent the transient gap, which causes a suppression in $\int_{\omega_1}^{\omega_2}\Delta\sigma_1(\omega)d\omega$, and the solid squares represent the transient Drude response, which lifts up $\int_{\omega_1}^{\omega_2}\Delta\sigma_1(\omega)d\omega$ to positive values in the same frequency region. It is clear that the transient gap and the Drude component develop with different time constants. The spectral weight suppression at \SI{40}{\milli\eV} reaches maximum at \SI{0.5}{\pico\second}, when the Drude component from photoexcited carriers is still on the rise. At around \SI{0.8}{\pico\second}, the transient gap is significantly weakened, and the Drude component reaches maximum. The transient gap decays with a shorter lifetime (\SI{0.19}{\pico\second}) than that of the Drude component (\SI{0.4}{\pico\second}). The existence of two time scales is consistent with previous pump-probe studies\cite{Rettig2012,Pogrebna2014}. Different time scales of the transient gap and Drude component suggest that they are separated non-equilibrium processes from different bands.

We now focus on the early-time optical response to investigate the temperature dependence of the transient gap. Figure \ref{fig:Ds1Tdep} presents $\Delta\sigma_1(\omega)$ from 0 to \SI{1.8}{\pico\second} at three temperatures. The transient gap formation is seen from below to above $T_{SDW}$ and persists up to room temperature. The transient gap develops at nearly the same energy for all temperatures, with weakened features at higher temperatures. The lifetime of the transient gap remains the same for below and above $T_{SDW}$ (see Supplementary Information).

\begin{figure} 
\includegraphics[scale = 1.0]{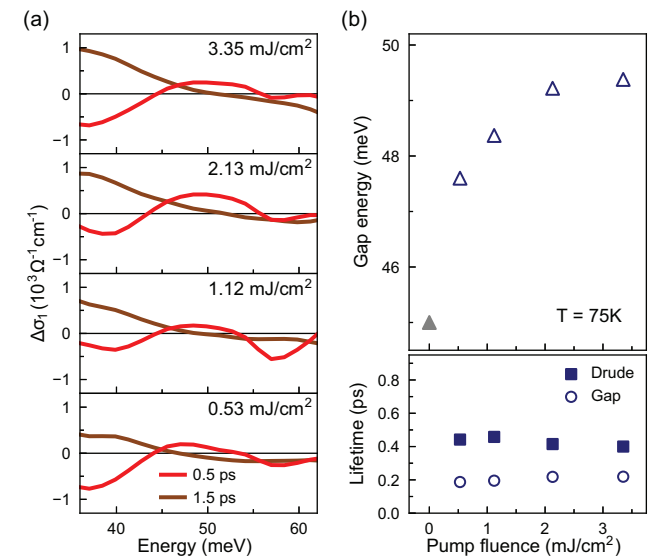}
\caption{\label{fig:Fluencedep} Pump fluence dependence of the transient state at $T = \SI{75}{\kelvin}$. (a) $\Delta\sigma_1(\omega)$ at two characteristic delays for various pump fluences. (b) With increasing pump fluences, the transient gap shifts to higher energies (upper panel), while the time constants of the transient state remain nearly the same (lower panel).}
\end{figure}

The pump fluence dependence of the transient state is shown in Figure \ref{fig:Fluencedep}(a). Using the peak position in $\sigma_1(\omega)$ to define the gap energy, we obtained the small SDW gap size as \SI{45}{\milli\eV} (Fig. \ref{fig:s1Ds1}(b)). Similarly, we identified the transient gap size from Fig. \ref{fig:Fluencedep}(a) using the peak position in $\Delta\sigma_1(\omega)$. At \SI{0.5}{\pico\second} time delay, a transient gap develops near \SI{48}{\milli\eV} with a \SI{0.53}{\milli\joule\per\square\centi\metre} pump fluence. The gap moves to higher energies with increasing fluences. When the pump fluence is approaching \SI{3}{\milli\joule\per\square\centi\metre}, the gap energy saturates (Fig. \ref{fig:Fluencedep}(b) upper panel). With the same procedure used for Fig. \ref{fig:s1Ds1}(d), we analyzed the time evolution of $\Delta\sigma_1(\omega)$ for each pump fluence. The extracted time constants are shown in the lower panel of Fig. \ref{fig:Fluencedep}(b). The lifetimes of the transient gap and the Drude component remain nearly unchanged with increasing pump fluences.

\begin{figure}
\includegraphics[scale = 1.0]{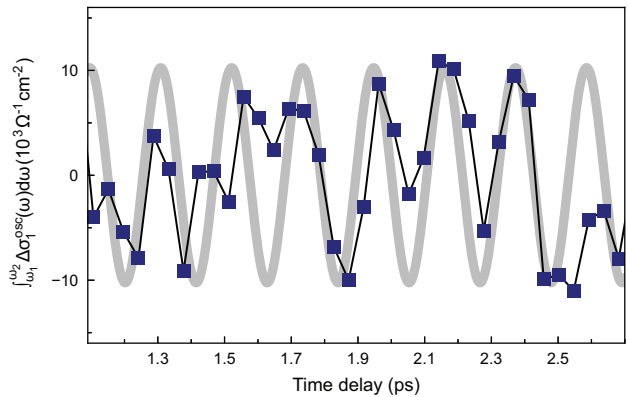}
\caption{\label{fig:osc} The phonon driven state evidenced by the oscillatory response in the optical conductivity. The navy squares are the integrated $\int_{\omega_1}^{\omega_2}\Delta\sigma_1(\omega)d\omega$, with the displacive component subtracted. Here, $\omega_1$ = \SI{12}{\milli\eV} and $\omega_2$ = \SI{20}{\milli\eV}. The data were taken at $T = \SI{120}{\kelvin}$ with a pump fluence of \SI{0.53}{\milli\joule\per\square\centi\metre}. A sinusoidal function with a \SI{5}{\tera\hertz} oscillation frequency (grey) is shown as a guide to the eye.}
\end{figure}

So far we have demonstrated a displacive response of optical conductivity in the phonon driven state. We now present the oscillatory response. Since the electronic band structure oscillates at the frequency of the driven phonon\cite{Yang2014,Okazaki2018}, the same oscillation should show up in the spectral weight of optical conductivity, as $\omega\sigma_1(\omega)$ is proportional to the joint density of states\cite{Dressel}. However, it is challenging to spectrally resolve the oscillatory response: it is a weak modulation on top of the large background displacive response. In addition, the signal-to-noise and time resolution in the terahertz region are orders of magnitude worse than that in the near-infrared region\cite{Kim2012}. To minimize the effect from the strong displacive response, fine time delay scans with a \SI{2.5}{\pico\second} time window were carried out for the low frequency region, where the signal-to-noise allows a qualitative analysis for the oscillation component. Figure \ref{fig:osc} shows the time evolution of the integrated transient conductivity change, $\int_{\omega_1}^{\omega_2}\Delta\sigma_1(\omega)d\omega$, with the displacive response subtracted. The spectral weight integral was evaluated with $\omega_1$ = \SI{12}{\milli\eV} and $\omega_2$ = \SI{20}{\milli\eV}. The grey line is an sinusoidal oscillation at \SI{5}{\tera\hertz}, serving as a guide to the eye. The oscillation is better resolved at later delays, when the transient gap disappears. Considering the limited time resolution of the THz probe, the oscillation frequency agrees qualitatively with the $A_{1g}$ phonon frequency probed by Raman spectroscopy\cite{Rahlenbeck2009,Chauviere2011}, and is consistent with the oscillation observed by previous pump-probe studies\cite{Mansart2010,Kim2012,Avigo2013,Yang2014,Gerber2015,Okazaki2018,Lee2022}. This verifies that the transient state we studied is a $A_{1g}$ phonon driven state.

We now discuss the possible origin of the light-induced gap at \SI{50}{\milli\eV}. A time-resolved photoemission study of BaFe$_2$As$_2$ reported a displacive downwards shifting of the chemical potential by \SI{50}{\milli\eV} at temperatures both below and above $T_{SDW}$\cite{Yang2014}. When the chemical potential shifts to lower energies, the low-energy bands which are occupied at equilibrium will become unoccupied; therefore, this may allow additional interband transitions in the transient state. For BaFe$_2$As$_2$, there are two $d_{xy}$ bands close to the Fermi energy. At the Brillouin zone boundary, the separation between the upper and lower $d_{xy}$ band is \SI{50}{\milli\eV}\cite{Pfau2019}. In the transient state, a \SI{50}{\milli\eV} downwards shifting of the chemical potential will allow interband transition between these $d_{xy}$ bands. Such an interband transition would be significantly shorter-lived than the chemical potential shift (\SI{0.13}{\pico\second}): as soon as the chemical potential shifts back from the maximum transient value, the available empty states at the upper $d_{xy}$ band will disappear. Since the transient gap observed here has a comparable lifetime to that of the chemical potential shift, it is not from interband transitions made possible by the chemical potential shift.

The transient gap is likely from a displacive band reconstruction as a consequence of a transiently increased pnictogen height. Time-resolved x-ray diffraction studies observed a displacive increase of the Fe-As distance in the phonon driven state. The maximum transient increase of the pnictogen height, considering the sum of the displacive and oscillatory components, is more than 5\% of the equilibrium pnictogen height with a pump fluence in the order of \SI{3}{\milli\joule\per\square\centi\metre}\cite{Gerber2015,Rettig2015}. As the electronic structure and the magnetic properties of iron pnictides are highly sensitive to the pnictogen height\cite{Gerber2015,Yin2011, Zhang2014}, a significant structural modification would visibly affect the SDW order. With increasing pump fluence, the pnictogen height increases linearly at low fluences and saturates at a fluence of \SI{3.5}{\milli\joule\per\square\centi\metre}\cite{Rettig2015}, which explains the fluence dependence of the transient gap in Fig. \ref{fig:Fluencedep}b. The lifetimes of both the displaced pnictogen height\cite{Rettig2015} and the driven $A_{1g}$ phonon\cite{Yang2014} show no systematic variation with pump fluences, which agrees with the fluence independent lifetime of the transient gap. Below $T_{SDW}$, since the transient gap takes the spectral weight from the equilibrium SDW gap (Fig. \ref{fig:s1Ds1}(c)), they possibly share the same origin: the transient gap is a blue-shifted SDW gap. This is consistent with the previous observations that an increased pnictogen height favors an enhanced Fe magnetic moment and an enhanced SDW transition temperature\cite{Gerber2015,Rettig2015}. Note that the transient gap develops and decays with different time scales than that of the transient Drude component (Fig. \ref{fig:s1Ds1} d), which can be understood as the following: the transient Drude comes from the thermalization of photoexcited carriers, which is commonly seen for metals\cite{Averitt2002}. It is a separate process from the transient modification of the lattice structure and band structure\cite{Yang2014,Rettig2015}, which lead to the transient gap. The photoexcited carriers and the transient gap coexist, since iron pnictides are multi-band materials. Finally, the appearance of the transient gap above $T_{SDW}$ is likely due to a light-stabilized SDW fluctuation in the normal state.

To summarize, we observed a transient gap generation at \SI{50}{\milli\eV} in BaFe$_2$As$_2$ when the $A_{1g}$ phonon is excited by laser pulses. The transient gap formation persists up to room temperature, indicating a robust band structure modification in the phonon driven state, which is likely from an enhanced SDW order as a direct consequence of a transient increasing in the pnictogen height. Our observation opens up new possibilities to investigate the interplay between the spin-density wave order, superconductivity, and nematic ordering in iron-based superconductors.

\begin{acknowledgments}
We thank N. L. Wang, R. Fernandes, and A. Klein for critical comments on this manuscript. This material is based upon work supported by the National Science Foundation under Grant No. 1944957. D. A. B. L. was supported by the US Department of Energy under Award Number DE-SC-0021305. Q. D. and G. C. were supported by the National Natural Science Foundation of China (Grant Nos. 11874417, 12274440), the Strategic Priority Research Program (B) of Chinese Academy of Sciences (Grant No. XDB33010100) and the Ministry of Science and Technology of China (Grant No. 2022YFA1403903).
\end{acknowledgments}

\bibliography{BaFe2As2.bib}% Produces the bibliography via BibTeX.

\end{document}